# Friendly thoughts on thoughtful friendliness

Adrian Kent [1,2]

[1] Centre for Quantum Information and Foundations, DAMTP, Centre for Mathematical Sciences,

University of Cambridge, Wilberforce Road, Cambridge, CB3 0WA, U.K.

[2] Perimeter Institute for Theoretical Physics, 31 Caroline Street North, Waterloo, ON N2L 2Y5, Canada

**Abstract**

We discuss Wiseman, Cavalcanti and Rieffel's "thoughtful" local friendliness no-go theorem and the experimental programme they propose to test local friendliness inequalities.   We argue that, to prove the theorem, the assumptions need to be strengthened to exclude the possibility of variable numbers of thoughtful agents existing in different phases of the experiment.   We argue further that this possibility may arise naturally, even in one-world versions of quantum theory.   We also query whether the motivations they give for their assumptions hold up well under their definition of "thoughtfulness" as displaying human-level cognitive ability, and suggest that their justification requires replacing "thoughtfulness" by "consciousness" or "conscious thoughtfulness".

0.  Introduction

There has been a sequence of intriguing proposals [Frauchiger-Renner 2018, Brukner 2018, Bong et al. 2020, Haddara-Cavalcanti 2022, Wiseman et al. 2022] for extensions of the Wigner's Friend thought experiment, including scenarios in which quantum predictions violate inequalities derived from metaphysical assumptions weaker than those used for deriving Bell inequalities.    All these proposals postulate one or more physical systems ("Friends") that, roughly speaking, are supposed to behave as an observer in essentially the same way as a human ("Wigner").   As in Wigner's original thought experiment [Wigner, 1961], it is supposed that each Friend can be isolated within a laboratory, so that from Wigner's perspective everything in the laboratory, including the Friend, may be treated as a closed quantum system.   The Friend may carry out observations of a microscopic quantum system within the laboratory.   Wigner can reverse the relevant quantum interaction before carrying out a measurement.   Thus Wigner is effectively a super-observer, able to carry out observations in complementary bases on the quantum state (which is taken to be to good approximation pure) describing the laboratory's contents.   The metaphysical assumptions are intended, inter alia, to define precisely what it means and implies to say that the Friends "behave as an observer in essentially the same way as a human." They also include a locality postulate or some related assumption.

Wiseman, Cavalcanti and Rieffel (WCR) recently [Wiseman et al., 2022] set out what they call a "thoughtful" local friendliness no-go theorem, which is intended to set out what would be required for future experimental tests of their "local friendliness" inequalities. It is also intended to motivate working towards such tests since, WCR argue, their metaphysical assumptions are widely held. WCR give quantitative estimates of the quantum computing resources required for a full test. These are far beyond current technology but may eventually be available. They also discuss intermediate milestones that can be motivated by stronger assumptions that might nonetheless perhaps still be relatively widely held.

WCR's paper is itself very thoughtful and friendly, in the sense that it aims to accommodate most stances on quantum theory and on the relationship between mind and physics and aims to treat these stances as respectfully and charitably as possible. WCR themselves avoid taking stances on the plausibility of different interpretations or versions of quantum theory, focussing on identifying which of their metaphysical assumptions is violated by each well-known version of quantum theory. They also aim to motivate and prove an interesting no-go theorem without taking a precise stance on the mind-matter relationship, by avoiding reference to consciousness in their metaphysical assumptions.

It is interesting and timely to examine just how far WCR succeed in their aims. Indeed, WCR have been actively inviting such scrutiny in order to build a community focussed on the motivations and possibilities for a long-term experimental programme testing their inequalities, using assumptions that could become weaker and increasingly plausible over time as technology advances.

This note questions whether WCR's metaphysical assumptions are well-defined in their present formulation, particularly when they involve "thought" and "absoluteness of events", and whether – even if we accept the definitions for the sake of discussion – they suffice to allow the derivation of their no-go theorem.

We also examine the plausibility of the assumptions. Logically, of course, this is irrelevant to the validity of the theorem: one can prove anything conditioned on a false assumption. Indeed, since the no-go theorem shows that quantum theory violates "local friendliness inequalities" that follow from WCR's assumptions, anyone who expects quantum theory to prevail in extended Wigner's Friend experiments has to accept that at least one of the assumptions must be false.

That said, the interest in the theorem and in experimental tests of the inequalities clearly depends on the assumptions having some plausibility. In the extreme case in which everyone was completely certain that one of the assumptions is incorrect, there would be little motivation for WCR experiments testing thoughtful friendliness.

There is also a more nuanced point. WCR's local friendliness inequalities are related to Bell inequalities, but – WCR argue – rely on strictly weaker assumptions than those needed to prove Bell inequalities. If the assumptions were logically strictly weaker, but there was no plausible way in which nature could exploit the logical gap, then there would be little motivation for carrying out thoughtful friendliness experiments rather than much simpler Bell experiments. This, of course, also applies to novel Bell experiments aimed at closing the collapse locality loophole [Kent 2005, Kent 2020] or any other remaining loophole in Bell tests. If any relevant loophole can be closed in both Bell and WCR experiments, and there is no plausible way in which nature can violate Bell inequalities and respect local friendliness inequalities in the respective experiments, then closing the loophole in Bell experiments should suffice.

Of course, plausibility and motivation depend on one's scientific worldview and priors.   No one can resolve these questions definitively for the whole scientific community, as the continuing lively debate over past and proposed Bell experiments shows.   We can, though, clarify our (collective) thinking, and perhaps reweight our priors, on extended Wigner's Friend experiments by exploring the implications of each assumption and its logical relationship to other propositions – as WCR do by, inter alia, identifying which approaches to quantum theory contradict which assumptions.    Our aim here is to contribute to this project.

### 1.  Ambiguities in Ego Absolutism

#### 1.1 WCR on Ego Absolutism

WCR's third metaphysical assumption is

> "Ego Absolutism. My communicable thoughts are *absolutely* real."

They explain this thus (numbers in quotations refer to footnotes and references in WCR's paper [Wiseman et al, 2022]):

> "Here you, the reader, when assessing whether you believe this assumption, should, of course, read it to yourself verbatim, not interpreting it as being a statement about the reality of the thoughts of me (the first author, as it happens)[9]. The formulation of the above assumption in the first-person singular is not frivolous. Like Wigner in 1961[10], I agree with Descartes [28] in the sense that I am more sure of the reality of my own thoughts than of anything else. Quite possibly the same is true for you, reader, again mutatis mutandis. This is, of course, does not mean that Ego Absolutism must be true. But if Descartes was on the right track, then Ego Absolutism is a philosophically minimalist criterion for an absolute ontology.
>
> The meaning of "absolute" in all of the above needs explanation (as flagged by its italicization in the definition of Ego Absolutism).  For my thoughts to be real in an absolute sense means that I do not have to qualify any statements about my thoughts as being relative to anyone or anything. For example, when I have a thought "it's alive!" it is not the case that this is my thought only in this world, or that this thought's existence is a fact only for me, or a fact only relative to certain other thoughts or states or facts or systems. Rather, my thought exists unconditionally. Or at least, by Ego Absolutism, this is the case for thoughts that are communicable rather than, for instance, ephemeral impressions or feelings whose existence I may even doubt myself."

This assumption does not hold in what WCR call "relativist" interpretations of quantum theory, by which they mean "interpretations that .. [say] my thoughts are not absolute, but rather relative".  They give as examples Fuchs and Schack's Qbism [Fuchs and Schack, 2013], Rovelli's Relational Quantum Mechanics [Rovelli, 1996] and Everett's "Relative State Interpretation" [Everett, 1957].   (They describe the latter as Everett's version of the Many Worlds Interpretation. I think it would be more accurate to call it Everett's attempt at a many-worlds interpretation, since it is now one of many such attempts, and there remains no consensus on whether it, or indeed any other, succeeds in producing a well-defined interpretation of quantum theory and has the properties and implications claimed.)

When combined with their Friendliness assumption, discussed below, Ego Absolutism is intended to ensure that we can assign a single definite value to a suitable Friend's observation in each run of the experimental test they consider.

## 1.2 Replicants

### 1.2.1 Replicants and consciousness

However, there is a lacuna in Ego Absolutism as framed.  This is best illustrated without assuming any specific interpretation of quantum theory, or even quantum theory itself.  Consider the sort of hypothetical scanning device discussed by Parfit [Parfit, 1984], which can create a perfect second physical copy of a human subject.   Suppose Bob is scanned and a replica created.   After the replication, we will refer to the scanned physical system as Bob0 and the replica as Bob1.

Now, we do not know that Bob1 will be conscious or have the same sense of continuing personal identity as Bob0.  We do not even know *for sure* that Bob1 will function in the world as a human similar to Bob0.   For example, dualists might suppose that the physical replication does not or need not necessarily replicate the mind, leaving Bob1 a zombie (in something close to the popular sense -- i.e., a mindless body unable to think and incapable of complex behaviours -- rather than the standard philosophical sense of a fully functioning human indistinguishable externally from Bob, but lacking consciousness).   In fact, we do not even know for sure that *either* Bob0 or Bob1 will function in the world as humans, in a way similar to that in which Bob functioned before the replication.   For example, dualists might also suppose that physical replication disrupts the mind-body connection, so that perhaps Bob0 and Bob1 are both zombies, or perhaps partial zombies with less than fully functioning minds.   Similar possibilities arise in some other views of consciousness that are not epiphenomenal or functionalist.   Another possible issue is that the hypothetical scanning device may contradict relevant physical laws.   For example, if Bob's consciousness and/or personal identity depend on his precise quantum state, the no-cloning theorem would preclude replication.

These uncertainties and caveats are important.   Still, there are plausible views of consciousness and personal identity according to which Bob0 and Bob1 are both conscious, fully functioning, and equally strongly convinced that they are the person whose life before the scan was Bob's.    Since WCR's arguments are not based on any specific view of consciousness or personal identity, we will assume for the sake of discussion that this is the case: Bob0 and Bob1 are fully functioning conscious individuals who regard themselves as continuations of Bob.   To simplify our discussion, we assume that Bob0 and Bob1 are aware of each other's existence but do not directly interact.

So, by assumption, Bob0 and Bob1 are independent fully functioning conscious humans. Immediately after the replication, they are in separate locations. They subsequently follow different paths and presumably have different experiences. So, their thoughts and memories also become distinct. All these reasons make it natural for us to discuss them separately: we can discuss where Bob0 is travelling, what he may be thinking, and how he is behaving, without any reference to Bob1. Is it reasonable then to say that Bob0's thoughts are absolute, and Bob1's thoughts equally absolute, in WCR's sense, as we normally would for a pair of distinct humans? Or are we supposed to say that there is still only one person, "Bob" – where the scare quotes are to remind us that "Bob" is unusual in now having two separate bodies -- and that, rather than referring to Bob0's or Bob1's thoughts, both we and "Bob" should refer to thoughts that "Bob" has relative to Bob0's or Bob1's body, respectively?

I follow Parfit in finding the first option natural, and the second extremely strained. Although Bob0 and Bob1 are identical immediately after the replication, their subsequent experiences distinguish them, and those experiences may – and presumably generally will – cause at least subtle differences in their characters and subsequent behaviours. Even if their post-replication experience goes no further than, say, observing 0 or 1 in otherwise identical experiments, we should expect their moods and subsequent behaviour to be subtly influenced by their distinct memories and associations of past experiences of 0 and 1. (WCR make essentially this point, in a different context, in their Appendix A. A similar point is also made in [Kent, 2021], discussing a thought experiment in [Chalmers-McQueen, 2022].) They may be more similar than any two unreplicated humans, but after the replication they are immediately distinguishable by location, and will generally also become at least subtly distinguishable by character and behaviour. By assumption, they are both fully functioning humans, whose conscious experiences appear to them as a natural continuation of those they remember having before the replication. The straightforward description of the situation is that there was one person before the replication, and now there are two. If so, Bob0 and Bob1 can both correctly maintain Ego Absolutism.

### 1.2.2 Motivations for considering replicants in Wigner's Friend experiments

The problem here for WCR's analysis is that this highlights the logical possibility that the number of subjects in Wigner's Friend experiments may be variable. In particular, it may be consistent to suppose there is one Friend before the experiment starts but two (or in principle many) Friends after the observation of the system within the isolated laboratory. It may also – this might in principle depend on which observation Wigner chooses – be consistent to suppose there is again only one Friend after Wigner's observation.

It should be stressed here that at this point in the discussion we, like WCR, are not assuming the validity of any particular version of quantum theory, or even of quantum theory itself. Of course, some versions of (or attempts at) many-worlds interpretations of quantum theory do propose something like this sort of picture. WCR intend to exclude standard many-worlds pictures (quantum or otherwise) from the scope of their theorem by requiring (Ego Absolutism) that my thoughts should be absolutely real, not real relative to this world, and requiring (Friendliness) that the same should be true of others of similar or greater cognitive ability. So we should also stress that we are not invoking a many-worlds picture here. What we have in mind here is a *one world* picture in which there are nonetheless sometimes two or more Friends who make different observations, and in which it is plausible to suppose that each Friend is conscious of their own observation.

Another distinction worth noting is that standard discussions of many-worlds interpretations of quantum theory generally imply that, if an observer measures a quantum system, and the measurement has (according to Copenhagen quantum theory) two possible results, then the observer is duplicated. In the Wigner's Friend experiments discussed by WCR, according to standard ideas about many-worlds quantum theory, this is true both of the Friend (who observes a qubit), and of Wigner (who observes the joint system of his Friend and the qubit). So Wigner is also duplicated, on these accounts, if he carries out a measurement with two possible outcomes. However, if we do not assume the validity of many-worlds approaches, or of quantum theory, duplication of the Friend in a Wigner's Friend experiment need not necessarily imply even the possibility that Wigner may be duplicated in the same experiment. For the loophole we discuss below, we suppose that the Friend may be duplicated, but Wigner is not.

Without referring to quantum theory, perhaps the best one can say to motivate considering this possibility is that (i) WCR argue that plausible assumptions about the experiments yield local friendliness inequalities, (ii) however, we want to consider at least the possibility that these inequalities may be violated (otherwise there is no interest in discussing the experiments), (iii) if they are violated then something counter to standard classical intuitions must be happening in the experiments, (iv) replication is a logically and physically coherent hypothesis, consistent with classical physics, worth considering if we are looking to go beyond standard intuitions.

In fact, of course, our strongest reason for taking seriously the possibility that the local friendliness inequalities will be violated in WCR's extended Wigner's Friend experiments is that this is what several versions of quantum theory predict. So we may sensibly look at quantum theory for further motivation for considering replicant or other hypotheses in the context of these experiments. Even if we are sceptical about the universal validity or completeness of quantum theory, we might reasonably think that the formalism of quantum theory may give some hints about what may be happening within these experiments.

We can do this without taking a firm stance on either the correct version of quantum theory or the relationship of consciousness to physics. Both are deep questions on which there is no consensus. It may be that no currently proposed stance on either issue is correct. Nonetheless, we can generate some intuitions about Wigner's Friend experiments without committing to a stance, simply by looking at the equations of textbook quantum theory. From Wigner-the-experimenter's perspective, after the Friend's observation and before his, the Friend's wave function involves a superposition of two macroscopically distinct states corresponding to the two possible observations. After Wigner-the-experimenter's observation of the Friend's observation, if he applies the projection postulate, as textbook Copenhagen quantum theory suggests he should, the Friend's wave function has only one component. If Wigner-the-experimenter instead reverses the Friend's measurement interaction, then again the Friend's wave function has only one component. It may be turn out to be wrong, and it certainly isn't the only option, but it seems quite natural for Wigner to think that he **should** apply the projection postulate after his observation – that's what the textbook says – while he **should not** apply it after the Friend's observation, given that the Friend is a quantum system isolated from the environment. Of course, there is a tension between what seems natural for Wigner-the-experimenter and what seems natural for the Friend who is the subject of Wigner's experiment. Textbook Copenhagen quantum theory is ambiguous about where to impose the Heisenberg cut, and in this thought experiment both the obvious options seem problematic: that was Wigner's point [Wigner, 1961]. So, a natural seeming perspective for Wigner-the-experimenter may turn out to be wrong. Still, it seems reasonable to consider it as one possible source of intuition. So we should be willing to consider Wigner-the-experimenter's perspective as suggesting we might hypothesize an underlying reality in which the experiment initially involves one Friend, there are two Friends with independent consciousnesses after the Friend's observation, and one single conscious Friend after Wigner's observation.

Again, we stress there are other interpretations of what is happening in the experiment available for Wigner-the-experimenter. There are also other interpretations available for the Friend, and indeed other interpretations may seem more natural from the Friend's perspective. Our point is simply that, before we get into detailed discussions of versions of quantum theory and of its relationship with consciousness, the picture just given ought to be on the table, along with others. **Some** description of the experiment must be correct; all the options (including perspectival pictures in which there is no observer-independent reality) are questionable; we see no a priori reason to exclude this one.

The reader might object that, even if there is no a priori reason to exclude this interpretation, we can rule it out once we consider the range of versions of quantum theory and stances on physics and consciousness currently on the table, since it is arguably not consistent with any of them.

The most obvious response to this is that all the known versions of quantum theory and all the standard stances on consciousness are problematic, and quite plausibly none of them is correct. The interpretation of the experiment just given might be justified by future developments even if it were not consistent with present ideas. In the context of the WCR theorem, this is enough reason to consider it.

I would add a second, perhaps more controversial, response. I think what we might call "Wigner's natural interpretation" **may** plausibly be consistent with, and even arguably natural within, some non-Everettian versions of quantum theory. Subsections 1.3 and 1.4 elaborate on this. (As noted in subsection 1.7, it may also arguably be natural within some Everettian versions of quantum theory that are not (or at least not explicitly) relativist.)

**1.2.3 The replicant loophole**

First, we explain why the possibility of multiple classical copies highlights a loophole in the WCR argument. We follow here the discussion and notation of their section 2.1, except that for notational convenience we relabel all experimental outcomes so that they belong to $\{0,1\}$ instead of $\{\pm 1\}$. WCR consider an extended Wigner's Friend experiment in which there are three observers, Alice, Bob and Charlie. Bob and Charlie, who are in separate laboratories, share a bipartite system. Charlie, who plays the role of Wigner's Friend, carries out a fixed measurement on his subsystem, with outcomes $c \in \{0,1\}$. Alice and Bob each have two measurement choices, labelled respectively by $x, y \in \{1,2\}$, with respective outcomes $a, b \in \{0,1\}$. They choose their measurements randomly, in regions outside the past light cones of the outcome $c$ and of the other party's choice.

If $x = 1$, Alice asks Charlie for the value of $c$, and takes the value given as her outcome, i.e., sets $a = c$. If $x = 2$, Alice carries out a different measurement on the contents of Charlie's laboratory, which include Charlie and Charlie's subsystem.

Now suppose that, in the interval between Charlie's measurement and Alice's, there are two copies of Charlie in the laboratory, $C_0$, who observes c=0, and $C_1$, who observes c=1. We suppose this is true deterministically, regardless of the choices of $x$ and $y$.

To disambiguate the notation, we refer to $C_0$'s outcome as $c_0$ and $C_1$'s as $c_1$, where $C_0$ is *defined* as the copy who observes $c = 0$, so that $c_0 = 0$ deterministically, and similarly $c_1 = 1$ deterministically. The probabilities for all other values of $(c_0, c_1)$ are zero. Thus we have

$$p(c_0 = 0, c_1 = 1 \mid x, y) = p(c_0 = 0, c_1 = 1) = 1 .$$

If x=1, Alice obtains an outcome a. From this she can infer that one of the Ci's observed the outcome $c_i = a$ .

But, if she understands the hypothesized ontology, she already knows that this must have been true, regardless of the value of a, and also regardless of the value of x. She knows before the experiment starts that there will be both a $C_0$ who observes 0 and a $C_1$ who observes 1. No measurement choice or outcome gives her new information about any absolute physical events within Charlie's lab.

So, the only nonzero joint outcome probabilities are

$$p(a, b, c_0 = 0, c_1 = 1 | x, y) = p(a, b | x, y) .$$

We no longer have a condition of the form given by WCR (on page 7):

$$P(a | c, x = 1, y) = \delta(a, c) .$$

But, without some such condition, we cannot run their argument. The only remaining constraint on $p(a, b|x, y)$ comes from WCR's Local Agency condition, which implies only that the correlations are non-signalling. Quantum correlations are non-signalling, so – if this loophole applies – there is no contradiction between WCR's assumptions (as stated in [Wiseman et al. 2022]) and quantum theory.

It is true that, if Alice understands the ontology, she might justifiably say, in runs where $x = 1$ and she observes $a$, that her measurement involved a communication with only one copy, $C_a$, and that her observation effectively destroyed the other copy $C_{\bar{a}}$ (where $\bar{a} = 1 - a$ ). Her measurement implies not just that one of the $C_i$ observed the outcome $c_i = a$ , but also that this $C_i$ is naturally identified with the copy she is now communicating with.
Still, this does not allow her to infer any new information about earlier absolute events from the postulates. In particular, she cannot infer from the WCR postulates that there was an earlier absolute event identifying the copy $C_a$ with whom she will later communicate.

In contrast, in WCR's analysis, which assumes a unique C, Alice can infer from their postulates that there was an earlier absolute event directly correlated with her observation, namely C's observation $c = a$ .

To reiterate: if the loophole applies, since $c_0 = 0$ and $c_1 = 1$ then hold deterministically, the only correlations to be explained are the joint outcome probabilities $p(a, b|x, y)$. Effectively, Alice and Bob are carrying out a Bell experiment on a system that includes a macroscopic subsystem. When $x = 1$, the identity of the $C_a$ who survives Alice's measurement determines her measurement result $a$, and neither the identity of this $C_a$ nor the value of $a$ need be identifiable in the ontology until the point where Alice carries out her measurement. When $x \neq 1$, Alice carries out some other measurement on the contents of Charlie's laboratory (including Charlie).

In the specific experiments WCR discuss in which quantum theory violates their local friendliness inequalities, if $x \neq 1$ Alice first reverses (what unitary quantum theory implies is) the unitary interaction between Charlie and the qubit, and then carries out some other measurement on the qubit. If unitary quantum theory correctly describes this part of the experiment, this effectively is a measurement on Charlie's subsystem alone, since Charlie's states at the start and end of the experiment are identical.

In any case, whatever measurement she chooses for $x \neq 1$, Alice's outcome $a$ again need not be identifiable in the ontology until her measurement. Bob's outcome $b$ is the outcome of a measurement on his subsystem, and of course also need not be identifiable in the ontology until the point where Bob measures. Quantum theory predicts the joint outcome probabilities $p(a, b | x, y)$, and any logically possible alternative explanation of these correlations would also suffice as a logically possible explanation of the outcomes of the WCR experiment. Of course, confirming the quantum predictions would confirm the violation of Bell inequalities, but Bell experiments already do that.

This has two implications. First, to imply the local friendliness inequalities, WCR's Ego Absolutism assumption needs to be adjusted to rule out the possibility of replicants. We discuss this further in subsection 1.5. Second, to the extent that one has credence that nature may produce replicants in extended Wigner's Friend experiments (EWFEs), one loses credence that these experiments would test anything more than Bell experiments do (albeit in a new regime). Probably few, if any, would dogmatically hold that replicants are being created, so this certainly does not remove all motivation for EWFEs. Perhaps many would start from the position that replicants are very unlikely and that the loophole barely diminishes the motivation for EWFEs. Against this view, we argue in the next two subsections that the replicant hypothesis merits significant credence.

We have focussed on the simplest extended Wigner's Friend experiment (EWFE), considered by WCR. The same essential point applies in more elaborate experiments in which Alice and Bob both carry out experiments on isolated laboratories that include Friends (respectively, Charlie and Debbie), who themselves carry out given measurements on a given quantum system within their respective laboratories. Alice and Bob may have more than two measurement choices, and Charlie's and Debbie's measurements may have more than two outcomes. For example, if Charlie's measurements give rise to M replicant Charlies, we can label them $C_0, C_1, \ldots, C_{M-1}$, where $C_i$ observed outcome $i$; similarly, we can label N replicant Debbies $D_0, D_1, \ldots, D_{N-1}$. The only nonzero joint outcome probabilities are

$$p(a, b, c_0 = 0, \ldots, c_{M-1} = (M-1), d_0 = 0, \ldots, d_{N-1} = (N-1) | x, y) = p(a, b | x, y)$$

and again WCR's postulates constrain these only by the no-signalling condition.

### 1.3 Replicant issues in one world versions of quantum theory

We have not so far assumed the validity of any version of quantum theory. However, we now argue that the possibility of replicants does in fact arise naturally in some "one-world" versions of quantum theory. For example, a mass density ontology defined by late-time electromagnetic field measurements [Kent, 2017] or by dynamical collapse models (see e.g. [Ghirardi et al., 1986; Allori et al., 2021]) might give a picture of two "half-density" versions of Wigner's Friend, classically superposed, during the experiment. Now it is not clear what we should hypothesize about how or whether conscious thoughts are attached to this classical superposition. More generally, we would argue, it is not a priori clear that there is a natural preferred hypothesis about how or whether conscious thoughts are attached to general physical states in any classical theory or any version of quantum theory. There are competing theories of consciousness even within classical physics, there are competing ontological proposals for most versions of quantum theory, and there are also competing proposals on the relationship of consciousness to any given quantum ontology. So, we cannot reasonably say that any given proposal is necessarily correct -- but nor should we exclude hypotheses that seem coherent and relatively simple. I would argue that, on the face of it, it is coherent and not particularly implausible to imagine that, in an ontology with two classically superposed "half density" friends, the Friend has effectively split into two Friends with separate consciousnesses, related to their precursor in the same sort of way that Bob0 and Bob1 are related to Bob in a Parfitian replication. It also seems coherent and plausible to imagine that there is again just one consciousness after Wigner's measurement, which replaces the classical superposition by a single "full density" Friend state. Again, Ego Absolutism does not exclude this, and nothing else in WCR's discussion covers this possibility.

Interpreting classical superpositions in a mass density ontology as representing multiple consciousnesses needs some discussion. The intuition is perhaps clearest when the classical superposition (and the underlying quantum superposition) is of two spatially separated, non-overlapping "half density" friends. In the standard understanding of mass density ontologies, a mass distribution corresponding to that of a single "full density" person simply *is* a person – that is what it means to say that the ontology is defined by mass distributions. And it is standard to assume that we can apply some (in detail unknown) theory of consciousness to the ontology, which tells us that (a) nothing not present in the ontology can be conscious, (b) full density people appearing in the ontology are (potentially) conscious in the way we are: normally conscious when behaving as awake and functioning, less conscious or unconscious when behaving as asleep and not dreaming, and so on.

Now there is nothing evidently important about the specific density of human bodies in this account. If a human behaving normally, but with density one-half or one-tenth of normal, appears in the ontology, it seems quite plausible that the relevant theory of consciousness would assign them the same conscious states as their full density counterparts. Of course, it is logically possible that a theory of consciousness could be density-dependent, but it would be somewhat surprising: most intuitions are that the state of a physical system's consciousness relates somehow to its information processing, not to details of its physical composition, and should not be sensitive to uniformly rescaling its mass density (at least by moderate factors -- see section 1.4 for further discussion).

If we are happy enough to suppose that a theory of consciousness implies that a single half density human has the same conscious state as her full density counterpart, we should be happy also to suppose that two non-overlapping half density humans each have conscious states, which are the same as those they would have if they were full density. And this seems reasonable even if those states are identical or nearly so. Again, there are certainly logically possible alternatives – but the supposition seems, at first sight, at least a plausible consequence of a theory of consciousness.

There are some similarities here with some Everettian ideas. A potential concern is that these similarities undermine the motivation for the combination of hypotheses considered, given that most proponents of collapse models and other one-world versions of quantum theories argue that Everettian ideas do not resolve the measurement/reality problem, while their preferred models can. So, it is worth emphasizing some important differences.

First, our discussion relies here on an explicit ontology, defined (in non-relativistic models) by mass density. While Everettians are similarly free to postulate a mass density ontology, they generally don't. Everettians generally argue that the unitarily evolving wave function already defines an adequate realist ontology (although Everettian accounts differ on how this is supposed to work, how the appearance of a quasiclassical world following standard quantum statistics emerges, and how (or even whether) we are supposed to be able to confirm the theory from our observations).

Second, our assignment of conscious mind states to physical states in the ontology does not involve probabilities that are proportional to the mass density. In fact, it does not involve probabilities at all. The suggestion, rather, is that, so long as the ontology contains a classical superposition of distinct macroscopic states of C with significant mass densities, there are two conscious C's, whether or not their densities are equal. This would be empirically problematic if the situation persisted for a long time in presently feasible experiments. However, it is compatible with the standard understanding of the models if (i) in regimes empirically tested to date, the ontological description quickly (compared to human perception time scales) becomes one in which only one has significant density, (ii) we can evade the so-called "problem of tails", discussed in the next subsection.

In contrast, Everettians need to explain the appearance of probabilities that are proportional to the Born weights of components of macroscopic superpositions. (One possible exception is a line of thought due to Leifer, who has considered a version of Everettian ideas in which Born weights play no direct role [Leifer, 2022]. As I understand it, this "ironic many-worlds interpretation" remains a work in progress. In any case, its aims and the structure of its ontology are very different from the one world proposal considered here.)

Third, whereas in typical Everettian models the physics of decoherence implies that branches subdivide but essentially never (at least in our cosmological era) recombine, here the two conscious replicas of C become one again by the end of the experiment WCR propose to demonstrate quantum violations of local friendliness inequalities, regardless of A's measurement choice.

**1.4 The problem of tails**

As flagged above, the discussion of the last subsection raises the "problem of tails". As stated so far, the argument given there also applies to a classical superposition of two humans with densities $\epsilon$ and $(1-\epsilon)$, for arbitrarily small positive $\epsilon$. In standard dynamical collapse models, applied to simplified descriptions of human experimenters observing quantum systems, the experimenter is described by this sort of classical superposition after the measurement, with a time-dependent parameter $\epsilon(t)$ that decays exponentially but never reaches zero. For dynamical collapse models, or other models that generate mass density ontologies with similar properties, to resolve the measurement problem, we need to conclude that the near-unity density human is real and conscious, and the near-zero density human is not real, or at least not conscious. So, our argument cannot apply for all $\epsilon$ in a one-world version of quantum theory if that version generally produces classical superpositions whose component densities remain non-zero for long times, and if we nonetheless want to maintain that said version resolves the measurement problem.

One could resort to postulating as a law of nature that consciousness is not present below some critical small density. But this seems a little desperate: we have just argued that a density-dependent theory of consciousness sensitive to uniform rescaling would be surprising.

Another option is to note that even the proponents of dynamical collapse models regard them as, at best, approximations to some underlying theory. One could imagine, or hope, that the relevant underlying theory implies that $\epsilon(t) = 0$ (precisely, not just to very good approximation), for all times $t$ shortly after a human-like observer looks at a quantum system. This might in fact be true of some versions of models in which the ontology is defined by asymptotically late time fictitious measurements of the electromagnetic or gravitational field [Kent, 2017].

A third option is to note that, in realistic models, there are very many microscopic contributions to the mass density ontology, arising not only from small and disappearing wave function components of macroscopic events (so to speak, from the "non-event" alternatives to the "event" that was actually realised), but also from wave function components corresponding to delocalized microscopic systems. As $\epsilon(t)$ gets smaller, these other contributions dominate. Looking at a plot of the mass density distribution, the "disfavoured" low-density human sinks over time into the background "noise". Plausibly, a theory of consciousness might imply that the disfavoured human's consciousness disappears as they fall beneath the surface of this microscopic sea, since the associated information processing and its underlying causal relationships no longer form a natural description of the mass density distribution.

We could try, speculatively, to make these intuitions more formal by imagining some sort of minimal description principle, in which the "natural" description of the evolving mass density is the one described by a minimal length program (or perhaps the program that minimizes some combination of length and run time). When the evolving mass density in a region corresponds, to very good approximation, to that of two non-negligible density humans functioning normally, one might think that the most elegant description involves describing each of them quasiclassically and summing them (together with other background contributions). This gives a preferred representation of the classical superposition, and we would then hypothesize that the theory of consciousness assigns consciousness (or not, depending on their brain state) separately to each component in this sum. On the other hand, when one human has density corresponding to a weight $(1 - \epsilon)$ and the other has density corresponding to weight $\epsilon$, where $\epsilon$ is astronomically small, one might think the most elegant description involves only the former added to background noise, since picking out the latter does not significantly simplify the noise contribution.

Finally, we need to consider classical superpositions of humans that overlap in position space. While a Wigner's Friend experiment involving a real human Friend could in principle be arranged so that the Friend moves to different locations depending on his observations, this would make the experiment harder. Relatively speaking, it would be easier to arrange this for a WF experiment involving a simulated intelligence (human or otherwise) on a quantum computer, but it still requires more programming steps and more qubits. So, the first experiments, of any type, are likely to involve Friends in superpositions of quantum states that have highly overlapping support in position space, and hence have classical mass density representations with the same feature. Nonetheless, the quantum states and their mass density representations would be distinguished by the different brain states (or, more realistically, simulated brain network states) corresponding to different observations. Plausibly, the most elegant description of the overall mass density would still be as a sum of the contributions from the Friends in different states (so long as they are in different states), together with background noise.

## 1.5 The need for a new postulate

The reader may perhaps find the possibilities described in 1.3 and 1.4 too counter-intuitive, or insufficiently developed to be persuasive. Even if so, the discussion in 1.2 stands. To prove a WCR-like theorem rigorously, one must exclude the possibility of variable numbers of subjects, not just be sceptical about whether it can be compatible with non-Everettian versions of quantum theory. So, something stronger than Ego Absolutism is needed. But what?

One option is to extend WCR's current postulates. For example, one might add:

UNIQUE EGO OVER TIME: At any future time $t$ there will be at most one thoughtful entity $E_t$ who identifies as a continuation of my present self. At any past time $t$ there was at most one thoughtful entity $E_t$ with whom I identify as a continuation. When $E_t$ and $E_{t'}$ are both defined, and $t > t'$, $E_t$ identifies as a continuation of $E_{t'}$.

UNIQUE EGO OF FRIENDS OVER TIME: If an independent party displays cognitive ability at least on par with my own, then UNIQUE EGO OVER TIME applies to them at all times.

However, if one also accepts the arguments of section 4, these postulates should be framed in terms of conscious thought, and can conveniently be combined into one:

UNIQUE CONTINUATION OF PERSONAL IDENTITY: if an observer $O_t$ is conscious at time t then at any time t'>t there is at most one conscious successor $O_{t'}$ who will identify with that observer and at any time t'<t there is at most one conscious predecessor $O_{t'}$ with whom that observer identifies. When $O_{t''}$ and $O_{t'}$ are both defined, and $t'' > t'$, $O_{t''}$ identifies as a continuation of $O_{t'}$.

Either way, the additional postulate(s) may well be less widely held than WCR's existing metaphysical postulates. The splitting of thoughtful or conscious entities as a result of something like Parfitian replication may or may not be possible, but it is not self-evidently implausible. The additional postulate(s) also seem rather pointedly tailored to rule out natural ways of thinking about Wigner's Friend experiments, which makes the result seem a little weaker. But if, as we argue, they (or something similar) are necessary, these costs must be accepted.

Similar comments apply to the assumption of Absoluteness of Observed Events in the earlier formulation [Bong et al., 2020].

## 1.6 Summary conclusions

The predictions of quantum theory are consistent with WCR's postulates as currently framed, since they do not rule out the possibility of replicated observers, each of whom can maintain the absoluteness of observed events. To establish a tight no-go theorem, in which the postulates imply inequalities violated by quantum theory, the postulates need to be extended. This is not hard, and we have suggested options.

The possibility of replicated observers of the type considered is more than a contrivance to make a logical point. It seems a relatively natural option *ab initio*, given that the sense of paradox in Wigner's Friend experiments comes from observers having apparently incompatible perspectives, and one possible resolution is that one perspective (Wigner's) is correct. We have argued that it also seems plausibly natural within some established approaches to quantum theory.

Of course, adding to or strengthening the postulates gives more scope for finding reasons for rejecting at least one of them, and if an EWFE does ultimately confirm the predictions of unitary quantum theory, while violating inequalities implied by a set of postulates, then at least one of them must fail to hold in the given EWFE. We think it is plausible that the predictions of quantum theory may be confirmed in an EWFE with suitably chosen Friends, that there are absolute events corresponding to the observations of these Friends within the given EWFE, and that a new/extended postulate along the lines we have discussed (for example, UNIQUE CONTINUATION OF PERSONAL IDENTITY) fails to hold.

(However, as we explain below, we believe these propositions are best discussed and most plausibly justified in the context of an empirically supported theory of consciousness; in particular, with an EWFE where the Friends consciously observe their measurement outcomes, according to said theory.)

### 1.7 Aside: do all Everettian versions of quantum theory require relative states?

As an aside, some modern Everettians might say that something like Parfitian splitting is also the correct way to understand what their version of many-worlds quantum theory says about the Friend's state after their measurement. They might argue that neither world nor branch are fundamental concepts and that it is thus not sensible to speak of a Friend's observations relative to their world or branch. Instead, they might say that in reality there simply are two (or more) Friends after the observation. On this view, this is not contradicted by the fact that the Friends are effectively unable to interact, and not directly aware of each other's existence: an absolute ontology can include non-interacting people or objects. WCR clearly intend Ego Absolutism to exclude this type of picture from the scope of their theorem, but, on our reading, as currently phrased, it does not. But, we stress, this is not our main point.

### 1.8 Other work

Müller and Jones [Müller-Jones 2022] have independently discussed the possibility of observer duplication in Everettian and other contexts as potentially relevant to discussion of Wigner's Friend experiments, and noted as an open question whether classical models might violate local friendliness inequalities.

## 2. Physicalism

In this section we consider WCR's second metaphysical assumption:

> "PHYSICALISM. Any thought supervenes upon some physical process in the brain (or other information-processing unit as appropriate) which can thus be located within a bounded region in space-time."

Our discussion in this section is meant to be independent of that in Section 1. It applies assuming WCR's original postulates but would also apply given new/extended postulates along the lines of those proposed in Section 1.

WCR comment on PHYSICALISM that:

> "This is a belief held almost universally by scientists. It is compatible with the monist assumption that a thought is nothing but a physical process, but does not require that. The reason for stressing the boundedness of the physical thought-process in space-time is that a physical process comprises a causally related set of physical events in space-time, as appearing in the first assumption, LOCAL AGENCY."

WCR may well be right that scientists almost universally hold this belief. Of the small minority who don't, most are probably dualists. But I think one need not be a dualist to believe there are interesting reasons to query whether it is necessarily true as stated.

The standard argument for the form of physicalism described by WCR as PHYSICALISM, as I understand it, runs something like this. We have a wealth of evidence that brain states and mind states (in particular, "thoughts", however these are defined) are correlated. Waking brain states, dreamless sleeping brain states and dreaming brain states are all easily distinguishable. Drugs that affect brain states also affect mind states, in ways that are well characterised and – in their high-level descriptions – correlated. Brain injuries affect mental capabilities, as do the effects of ageing on the brain. Brain imaging studies show detailed correlations not only between brain states and visual perceptions, but also between brain states and imagined visual images. And so on: a wide variety of data suggest that brain states strongly constrain mind states; the simplest hypothesis is that brain states determine mind states; no data contradict this. Perhaps the jury is out on whether the nervous system outside the brain also plays some role – but nothing beyond the brain and nervous system appears relevant to understanding mind states.

I don't query this account of the data, but there is at least a loophole in this interpretation. For reasons discussed further in later sections, I think it is more natural to frame the discussion in terms of "conscious thoughts and perceptions" rather than WCR's proposal, "thoughts", which seems to me less cleanly defined. So I will discuss physicalism as referring to consciousness. However, the discussion in this section is also intended to apply if framed in terms of "thoughts": either way, the issue is identifying the physical substrate.

Consider for example a simple visual perception – seeing a flash of light as a result of a few photons hitting the retina in an otherwise dark room. At least one retinal cell is excited.

This causes a signal in the optic nerve, which transmits to the visual processing regions in the brain, causing neurons to fire. Are we conscious of the flash because of the retinal cell excitation, or the first neuron to fire, or later steps in the visual processing, or something else? I don't think we know precisely. Perhaps the relationship between consciousness and material physics will turn out to be such that the question doesn't make sense or doesn't have an unambiguous answer. But it plausibly might. If, as for example the proponents of integrated information theory (IIT) [Tononi 2008, Oizumi et al. 2014, Tononi 2015] hope, some future theory of consciousness gives mathematical rules that ascribe conscious mind states to specific types of physical system in specific types of state (or sequences of states) then the theory (at least if it resembles IIT in this respect) should tell us at what point in the causal chain conscious perception is created. It might tell us, for example, that some specific step in the brain's visual processing evokes consciousness, so that if the processing is somehow halted before that step the subject has no conscious perception of the flash. This may be difficult to test directly but would still deserve some credence if implied by a sufficiently parsimonious theory with sufficient explanatory power.

Now, the physical causal chain continues beyond the behaviour of brain cells. Each excitation generates a perturbation of the electromagnetic field, which interacts with the brain and body. Some resulting photons may, perhaps after multiple scatterings, absorptions and re-emissions, propagate outside the head, perhaps even to future infinity. The induced motions of ions and molecules in the cells also perturb the gravitational field, albeit very slightly, since the densities of the relevant chemical species are not identical. Moreover, our perceptions often also produce physiological responses. Perhaps, indeed, they essentially always induce some physiological response, though this response may generally be very slight, involuntary and unconscious. These physiological responses too perturb the electromagnetic and gravitational fields, generally having a stronger effect on the latter than do the intra-cellular effects involved in brain processing.

All these perturbations could, plausibly, be faithfully correlated with the brain states, in the sense that in principle measuring very precisely the electromagnetic or gravitational field changes could allow one to determine the brain state. If so, these field states are equally correlated with the mind state. All the evidence for physicalism is -- potentially -- equally consistent with the hypothesis that the conscious mind state supervenes not on the brain itself but on the field states around the brain. It may also be consistent with the hypothesis that the conscious mind state may supervene on the subset of field states that propagate to future infinity. It may too be consistent with the hypothesis [Kent, 2017] that the conscious mind states we think of as associated with the brain at approximate position x at time t are correlated with the subset of field states that propagate to future infinity outside the future light cone of $(x, t)$.

It may at first sight seem unlikely, even fantastic, that consciousness is fundamentally associated with the after-effects of brain activity rather than with the activity itself. But the fact (if it is) that brain states determine conscious mind states is something that should ultimately follow from a theory of consciousness (which we do not presently have) rather than taken as dogma. Finding a theory of consciousness that implies that conscious mind states are indeed determined by brain states, and not by the associated states of the environment, is an explicit ambition of IIT, for example. In principle, IIT is intended to apply to causal chains throughout the physical environment. IIT proponents aim to develop axioms that start from the micro-level, do not refer to brains or living organisms or any other biological concept, but nonetheless imply that each human and animal brain normally has a separate consciousness that can normally be characterized without describing every detail of its environment. Whether or not IIT succeeds in this, one can imagine other theories of consciousness in which human and animal conscious mind states are associated with a different physical system. This physical system might include the brain and nervous system; relevant brain activity might be a necessary but insufficient condition for consciousness. Indeed, there are some speculative ideas attempting to extend IIT to fundamental fields (e.g. [Barrett, 2014]) and other theoretical (e.g. [McFadden 2020; Ward-Guevara, 2022]) and experimental (e.g. [Pinotsis-Miller, 2022]) work exploring the association of consciousness with electromagnetic fields.

As WCR note, specifying that thoughts supervene on the brain, or some other appropriate information processing unit, implies that the thoughts are located within a bounded region in space-time (at least if the supervenience is such that the thoughts occur during the information processing). Supervenience on the electromagnetic or gravitational fields, or other correlated physical systems, does not necessarily guarantee this. One might or might not take seriously any of the specific ideas mentioned above. Still, the issue that we need to assume something about the substrate for consciousness should at least be acknowledged. One could for example rename the assumption BRAIN PHYSICALISM, removing the mismatch between the current name (which makes no reference to the brain) and the current definition of PHYSICALISM (which does), or relax it slightly to BRAIN AND NERVOUS SYSTEM PHYSICALISM, defined in the obvious way. This still leaves some ambiguity about whether the term includes supervenience on electromagnetic or gravitational fields that are associated with the brain. In fact, the key point here is not precisely what is the substrate for the supervenience but where in space-time it is. The key point of the assumption is that thoughts supervene on material within a region in space-time, bounded in space by the head or (perhaps) body and in time by the interval during which we believe the relevant thoughts took place. So perhaps SPACE-TIME BOUNDED PHYSICALISM, appropriately defined, might best capture what is required.

(A general problem in testing ontological hypotheses is that this interval is hard to pin down precisely, especially since neuroscientific data suggest that subjective timings are somewhat unreliable. Making any of these physicalist assumptions precise thus relies on arguments about how unreliable our subjective timings could plausibly be. We hope to pursue this interesting issue in a broader context elsewhere. Here we assume that some reasonable estimate of when thoughts occur, accurate to within seconds, can be justified, at least as plausibly enough that EWFEs assuming this estimate are agreed to be well motivated.)

### 3. What is a "thought" if not a conscious thought?

This section considers WCR's discussion of "thoughts" and "thoughtfulness". The discussion applies assuming WCR's original postulates. It would also apply assuming any of the alterations or additions suggested in sections 1 and 2 (except where they pre-empt the points made here by referring to consciousness).

**3.1 WCR's discussion of thoughts**

In WCR's words, they

> "intentionally use the word "thoughts", rather than, as Wigner used, "consciousness", because the former seems easier to identify and less controversial — we wish to avoid debates about what constitutes consciousness or even whether it exists. The connotations of the word

"consciousness" may also make humans unwilling to ascribe it to a non-human intelligence. Similarly, we wish to avoid having to speculate as to whether a given intelligence has the "same types of impressions and sensations" [2] as our own."

This is reflected in some of their definitions, which we recap here:

"PHYSICALISM. Any thought supervenes upon some physical process in the brain (or other information-processing unit as appropriate) which can thus be located within a bounded region in space-time."

"EGO ABSOLUTISM. My communicable thoughts are *absolutely* real."

"FRIENDLINESS. If, by open-ended communication, an *independent* party displays cognitive ability at least on par with my own, then they have *thoughts*, and any thought they communicate is *as real as* any communicable thought of my own."

They add that:

"The italicizations indicate concepts that are elaborated in the text. In particular, "the italicization of … "thoughts" is so we (the authors) can take the opportunity to reëmphasize that "thoughts" is not to be equated with "conscious thoughts" or "qualia" or "sensations" or "experiences". We reject the equation because it may make the assumption of Friendliness less plausible, or less well defined, for some scientists or philosophers."

I am not persuaded that this move to dilute controversy and make FRIENDLINESS more user-friendly works. WCR are clear about what they do *not* wish to equate with "thoughts", but (on my reading) unclear about the precise intended meaning of "thoughts".

### 3.2 Considering conscious and unconscious thoughts

What *is* a "thought" if it is not necessarily a "conscious thought"? An obvious response is that a thought may be unconscious. If we take that to be oxymoronic, we are left equating "thought" and "conscious thought" after all, in which case WCR's move is ineffective. But it does not seem necessarily oxymoronic: I am happy to accept the idea of an "unconscious thought" as a proposition that is not internally voiced but is nonetheless registered in the brain and affects my overall worldview or behaviour. Similarly, an "unconscious perception" refers to some aspect of the environment that I'm not consciously aware of but that is unconsciously factored into my overall worldview or behaviour. Unconscious thoughts can be communicated: for example, observant and psychologically acute people may infer them indirectly from a subject's behaviour, even if he himself remains unaware of their influence. Unconscious perceptions can seemingly be communicated to others too, as experiments on "blindsight" famously suggest (see e.g. [Humphrey, 1974; Weiskrantz, 2009; Humphrey, 2022]).

If we accept these ideas and take "thoughts" to mean "conscious or unconscious thoughts", we get a working definition of a "thought" as something like "a proposition represented by some appropriate coding in the relevant information processing system" and a "communicable thought" as a thought that influences (or can influence) behaviour in a way that allows independent observers to infer the relevant proposition. These definitions indeed not only don't presuppose that thoughts (even communicable thoughts) are necessarily conscious, but also make sense even if one is uncertain what consciousness means. To this extent they satisfy WCR's aim to have definitions that are widely agreed to be well-defined.

The problem, though, is that these definitions make EGO ABSOLUTISM and FRIENDLINESS much less credible, at least when combined with some widely held stances on consciousness. I'm happy to agree that my communicable *conscious* thoughts are real, and in normal circumstances absolutely real in WCR's sense. "Cogito ergo sum" seems a good basis for our understanding of reality, if we translate "cogito" as "I consciously think", or better, "I consciously experience" (which covers both thoughts and perceptions). But I don't see any reason to give any similarly privileged ontological status to any communicable *unconscious* thoughts I may have. Describing a communicable unconscious thought is making a statement about some particular physical processes in the brain, which are presumably fundamentally represented as statements about its evolving quantum wave function, and which may or may not be represented in the underlying ontology. Descartes' principle doesn't apply to them, and it isn't evident that they must necessarily be real in any plausible ontology.

Admittedly, it's at least arguable that communicable unconscious thoughts might have some features that are not shared by generic physical processes. But it seems hard to identify these features in simple terms. For example, they are part of a computation – but so are many mechanical operations, and in some sense so are all physical interactions. And we can communicate them – but we can also communicate many things about the world beyond our brains.

Let me stress again that I don't deny the possibility that unconscious thoughts are real, or even absolutely real. For example, if the correct version of quantum theory turns out to imply that a fine-grained quasiclassical description of the dynamics of the matter comprising the brain is real, then indeed, there is a sense in which our unconscious thoughts are real – namely that they are useful high-level descriptions of interesting aspects of the physical ontology. But, so far as we know, there is nothing qualitatively distinctive about the matter in the brain, and WCR certainly do not want their argument to require that there is. If the correct version of quantum theory assigns reality to a quasiclassical description of the brain's dynamics, it presumably assigns reality to a quasiclassical description of the dynamics of many other (maybe all) systems of matter that have quasiclassical descriptions. There are certainly interesting versions of quantum theory in which high-level quasiclassical descriptions can be defined from the ontology. But these versions of quantum theory go far beyond the assumptions WCR – who, recall, do not even assume the validity of quantum theory -- wish to make. Indeed, WCR argue in their section 5 that the best known of these versions, Bohmian mechanics and dynamical collapse models, each contradict at least one of their assumptions.

I don't even mean to deny the possibility that conscious and unconscious thoughts are absolutely real, and essentially nothing else is. There might conceivably be an interesting ontology in which the relevant physical processes are real, while all other physical processes in the brain, body and environment are not. This seems close to framing idealism in realist language, and an obvious concern is that any idea along these lines would have all the problems of idealism without the metaphysical simplicity that gives it (to proponents) some appeal. If so, it seems an unattractive option. More importantly, even if some case were found for it, this proposal too would go far beyond WCR's assumptions.

In summary, there seems no principled reason, within WCR's framework, why we should ascribe reality, let alone absolute reality, to unconscious thoughts. But if we don't, then EGO ABSOLUTISM fails. If you (the reader) have a single communicable unconscious thought, you have a thought that you have no good reason to regard as absolutely real.

FRIENDLINESS also fails. Another party might have cognitive ability greater than or equal to mine, with their cognition working entirely through unconscious thoughts. If so, while it is correct to take their displayed cognitive ability as evidence that they have thoughts, it is wrong to infer that any thought they communicate is as real as my conscious communicable thoughts.

Even what one might call SELF-FRIENDLINESS fails: I have unconscious communicable thoughts as well as conscious communicable thoughts, and (on this view) the former are not as real as the latter.

None of these points affects the logical status of the theorem, which derives conclusions from postulates. But they do make it potentially less interesting: if the relevant postulates turn out to be less credible than they initially seem, or even obviously wrong, there is less motivation for testing them against quantum theory. WCR aim to find postulates that, inter alia, are widely held amongst physicists and other scientists. So it is worth fleshing out their implications and their consistency with standard stances on thought and on consciousness to establish how widely held they are, and whether there might be more widely held alternatives that still allow a useful no-go theorem.

### 3.3 Other options?

I struggle to find another option more helpfully aligned with WCR's discussion without referring directly to consciousness. One could simply deny the existence of consciousness -- but then "Cogito ergo sum" is no longer persuasive. Alternatively, one could frame the definition of "thought" to exclude unconscious thoughts – but then the argument effectively refers to "conscious thoughts", which is not WCR's intention.

Another option could perhaps be to deny that "conscious" can qualify "thought" in the binary way suggested above. Perhaps it is reasonable to speak of "partly/almost/barely conscious thoughts"; possibly it might even be reasonable to maintain that no thought is either fully conscious or fully unconscious. But these moves do not seem to help. On the standard understanding of ontology, it isn't reasonable to talk of "partly/almost/barely real" entities, so we must take a stance as to whether any given type of quasi-conscious thought is or is not real. And then we are back again with the possibilities already considered above: all thoughts are real, even completely unconscious thoughts (which is not supported by a Cartesian argument and hard to justify otherwise within WCR's framework); all thoughts are real because all thoughts are at least slightly conscious (which is still hard to justify by a Cartesian argument: I am more sure of the reality of my own thoughts than of anything else when they have the immediacy and clarity characteristic of vivid conscious experiences; I am not at all sure of the reality of thoughts that may lie so close to the borderline of unconsciousness that I cannot grasp or characterise them); some thoughts are real and some are not (so EGO ABSOLUTISM and FRIENDLINESS fail); no thoughts are real (eliminative materialism, which contradicts EGO ABSOLUTISM). Another issue with the second option is that by equating "thought" with "(at least slightly) conscious thought" it invokes consciousness, and uses this to justify the reality of thoughts via Descartes' principle, a line which WCR aim to avoid.

## 4. Biting the bullet on consciousness

In this section we consider the possibility of replacing WCR's EGO ABSOLUTISM assumption by CONSCIOUS EGO ABSOLUTISM, PHYSICALISM by CONSCIOUS PHYSICALISM, and FRIENDLINESS by CONSCIOUS FRIENDLINESS. The discussion here applies assuming that WCR's other postulate and assumptions hold. It would also apply assuming any of the alterations or additions suggested in section 1. As we note below, we can easily also incorporate the alterations suggested in section 2.

### 4.1 CONSCIOUS versions of the thoughtful assumptions

### 4.1.1 CONSCIOUS EGO ABSOLUTISM

We have argued that WCR's move to avoid explicit reference to consciousness makes it hard to justify EGO ABSOLUTISM and FRIENDLINESS, if "thought" is not synonymous with "conscious thought". In particular, it undercuts their Cartesian argument for EGO ABSOLUTISM. If so, to produce an interesting theorem (i.e., one whose premises are plausible), we need to replace "thought" by something like "conscious thought" in these two assumptions. Thus we are motivated to replace EGO ABSOLUTISM by

CONSCIOUS EGO ABSOLUTISM: My communicable conscious thoughts are absolutely real.

Those who think consciousness undefinable will obviously reject it. For those comfortable in principle with assumptions that refer to consciousness, the motivation is hopefully clear from our earlier discussion, even to those who prefer WCR's version.

### 4.1.2 CONSCIOUS PHYSICALISM

For consistency, we should also replace PHYSICALISM by

CONSCIOUS PHYSICALISM: Any conscious thought supervenes upon some physical process in the brain (or other information-processing unit as appropriate) which can thus be located within a bounded region in space-time.

Following the discussion of section 2, we suggest improving this by

CONSCIOUS SPACE-TIME BOUNDED PHYSICALISM: Any conscious thought supervenes upon some physical process in an appropriate bounded region in space-time.

Neither of these changes should significantly affect the discussion. The new postulates are logically implied by those they replace. While the converse is not true, we expect that almost everyone who accepts the CONSCIOUS versions would also accept the originals.

**4.1.3 CONSCIOUS FRIENDLINESS**

Redefining FRIENDLINESS is less straightforward. Before we address consciousness, we note that FRIENDLINESS, as framed by WCR, appears to assume that the Friend's reports are honest. We do not want to take this for granted. We can lie about our own thoughts and our own conscious experiences, so we need to allow that other parties can too. A human-level AI could, for example, be maliciously programmed so that, in response to almost all inputs, it engages as honestly as it can in human-level cognitive conversations, but whenever it's asked about the outcome of a measurement it carried out on a qubit, it gives a pseudo-random answer. Or it could misrepresent other categories of thought. To address this sort of possibility, we would suggest replacing FRIENDLINESS by

FRIENDLINESS': If, by open-ended communication, an independent party persuasively displays cognitive ability at least on par with my own and also persuasively displays honesty, then they have thoughts, and any thought they communicate is as real as any communicable thought of my own.

Here we require that the party must display honesty, in order for us to accept that they reliably communicate thoughts. The word "persuasively" is added (twice) to emphasize that we must be able to continue the communication, using whatever tests we deem appropriate, until we are persuaded both of its cognitive ability and its honesty.

To reframe this in terms of conscious thoughts, we propose

CONSCIOUS FRIENDLINESS: If, by open-ended communication, an independent party persuasively displays cognitive ability at least on par with my own and also persuasively displays honesty, then they have conscious thoughts, and anything they report as a conscious thought is as real as any communicable conscious thought of my own.

Together with CONSCIOUS EGO ABSOLUTISM, this implies that, if a Friend reports they were conscious of a particular measurement outcome, there was a real event corresponding to that observation. Together with CONSCIOUS (SPACE-TIME BOUNDED) PHYSICALISM it allows us to localize the real event, if we believe we can identify the appropriate information-processing unit (or bounded region in space-time).

**4.1.4**

WCR are clearly correct that some scientists and philosophers will find CONSCIOUS EGO ABSOLUTISM and CONSCIOUS FRIENDLINESS not well defined, simply because they refer to consciousness.[1] Some may also reasonably find the reference to persuasive evidence of honesty in CONSCIOUS FRIENDLINESS problematic: how exactly should we look for such evidence, and can it be truly persuasive? (The same question applies to persuasive evidence of human level cognitive ability, of course, but some may reasonably believe it is easier to answer in this case.)

---

[1] Some will also find the terms "intervention" and "real" in WCR's assumptions to be ill=defined. As WCR recognise, there probably is no universally agreed set of definitions that allows an interesting result along the lines of their no-go theorem. The realistic aim is to find one or more sets of definitions that are as widely agreed as possible.

Even among those happy enough with the definitions, or at least with their application in some interesting classes of extended Wigner's Friend experiment, many will probably find the assumptions not plausible. Still, we find it hard to see better alternatives, having argued that the existing assumptions are either not well-defined (if "thoughts" means something other than "conscious or unconscious thoughts") or implausible (neither EGO ABSOLUTISM nor FRIENDLINESS are plausible if "thoughts" means "conscious or unconscious thoughts").

On the other hand, many scientists and philosophers *do* think "consciousness" and "conscious thought" are well-defined terms. Some will find CONSCIOUS EGO ABSOLUTISM both well-defined and credible, and some of these may well also find a version of CONSCIOUS FRIENDLINESS well-defined and credible. Looking at the broader intellectual picture, there are several well-defined stances about the relationship between consciousness and physics. One can list those stances and identify which assumptions (if any) they reject, just as WCR already do for interpretations of quantum theory, and this seems a natural way to refine the discussion further.

### 4.2 An elegant theory of consciousness likely would not support CONSCIOUS FRIENDLINESS

I will do this here just for the stance I currently find most interesting, namely that there is some as yet unknown relatively elegant mathematical theory of consciousness that gives axioms and equations from which we can infer whether, to what extent, and (in some sense yet to be understood) in what way any given physical system is conscious. That is, a theory that achieves something like what IIT aspires to – although I doubt that, if there is a good theory to be found, it will resemble the present formulations of IIT.

#### 4.2.1 Problems with elegant mathematical theories of consciousness

There are serious difficulties with the idea of an elegant mathematical theory of consciousness. I still give it credence because every other stance on consciousness and physics seems at least equally problematic. I find it interesting because it seems natural from the perspective of modern ideas in the foundations of physics, particularly of quantum theory. A natural approach to the quantum reality problem is to look to extend quantum theory by beables, mathematically defined quantities that are correlated with the evolving wave function and, generally, with other variables in the theory, and are postulated to "represent reality". A mathematical solution to the quantum reality problem would allow us, in principle, to identify what, if anything, is real about a given physical system. Similarly (though much more ambitiously, given the current state of our understanding) we could hope for mathematical rules that identify what, if anything, a given physical system is conscious of.

This stance makes no commitment about the form of the relevant mathematical rules. They might relate to the details of information flow in networks, as IIT postulates. In this case, it's possible that two systems with the same functionality but different internal networks can have very different conscious states. For example, IIT implies that, for any conscious networks, one can construct a functionally equivalent unconscious network (see e.g. [Doerig et al.,2019; Tsuchiya et al., 2019]).

Alternatively, they could, in principle, depend on the physical composition of a system as well as its abstract description as an information processor. For example, though it seems a priori unlikely, in principle they could require conscious systems to have components within a certain mass range, or to contain certain types of particle or molecule. Somewhat more plausibly, they could depend on the details of a physical system's quantum state. For example, they could imply that a Wigner's Friend in nontrivial superposition is unconscious, or conscious in only one branch, or differently conscious from a Friend in an almost pure state.

Of course, these last possibilities seem implausible to those persuaded by some form of quantum functionalism (see e.g. [Wallace, 2012]). But functionalism is just another theory of consciousness, unsupported (though not contradicted) by empirical evidence. Even at the classical level, functionalism has serious problems (see e.g. [Levin, 2021]), and many physicists and philosophers of mind find it quite unpersuasive.

A broader concern is that axiomatic mathematical theories of consciousness tend to imply that two systems (for example, a human and a computer) can behave in exactly the same way in response to stimuli, and report exactly the same internal mind states and memories, with one of them accurately reporting its conscious states and the other not (for example because it has never been conscious). This, some (e.g. [Doerig et al., 2019]) argue, puts such theories beyond the realm of science: if a theory predicts that some system's reports of its conscious state will be consistently and radically misleading, we can't directly test these predictions, since the system's reports are the only relevant data we have.

### 4.2.2 In defence of elegance

Everyone agrees that a science of consciousness faces obstacles. A purportedly objective theory of subjective experience necessarily makes predictions that cannot be directly tested. Nonetheless, I agree with the proponents of IIT (e.g. [Tsuchiya et al. 2019]) that to say that such a theory is necessarily unscientific takes too rigid a view of science. Parsimony is a valid scientific criterion for preferring one theory over another, even when they make the same predictions. A theory that, using only simple axioms applied to brain imaging data, gave a description of human mind states is testable against self-reports. It's unlikely that any theory will agree with every self-report: people lie, or can be un-self-aware, or in states of psychotic delusion. But if typical humans, including you (the reader) agree that its description of their mind state is typically correct, I imagine you would agree the theory would be a major scientific achievement. If we had no other similarly elegant theory that replicates its success, we should be inclined to take the theory's predictions about non-human physical systems seriously: that is, we should give them significant credence (if perhaps less than we give its predictions for a so far untested human).

### 4.2.3 Summary

On this view, if we hold out hope for an elegant mathematical theory of consciousness, we are obliged to reserve judgement on CONSCIOUS FRIENDLINESS, since we may well end up giving significant credence to a theory or class of theories that contradicts it. In fact, CONSCIOUS FRIENDLINESS should be seen as just one possible assumption on which to base a proto-theory of consciousness. Ignoring for the moment the possibility of quantum superpositions, and considering effectively classical states, it says, inter alia, that any human-level AI is as conscious of its communicable thoughts as we are of ours. This arguably fits well with functionalism, which I think is also best understood as a speculative attempt at a proto-theory of consciousness. But we have no good reason to reject other theories that say that a human-level AI may be more conscious than us, or less, or not at all, depending on its architecture, or the underlying technology. Nor should we pre-judge the possibility that the rules associating consciousness with a quantum superposition state may not agree with either CONSCIOUS FRIENDLINESS and EGO ABSOLUTISM (as would be required to run a consciously friendly version of WCR's argument) or Everettian quantum functionalism (some informal version of which is probably the most common alternative intuition among physicists about the consciousness of Wigner's friend).

For these reasons, we argue, WCR's ambitious experimental programme cannot produce truly persuasive results unless and until we have an evidence-based theory of consciousness. However, it could produce results that are persuasive modulo some plausible, albeit speculative, stances on consciousness. It might possibly even produce results that advance us towards an evidence-based theory of consciousness. These seem sufficiently strong motivations for pursuing it.

5. **Other comments**

a) WCR's discussion of FRIENDLINESS suggests that their argument should run given any standard view of the mind-matter relationship. I am not sure this is true of eliminative materialists, who, as I understand their stance, must believe their thoughts are "not real at all", since the concept of "thoughts" has no referent in reality. It is not clear to me that they can sign up to every proposition entailed by FRIENDLINESS. They can agree that any thought communicated by an independent party is as real as any communicable thought of their own, since in both cases the thoughts are "not real at all". But can they then also agree that an independent party displaying cognitive ability on a par with theirs has thoughts? Perhaps, if they think that all thoughts are unconscious, and also not real. But if they take this line, it's not clear they can also agree with EGO ABSOLUTISM, which requires them to accept that their communicable thoughts are absolutely real, and hence a fortiori real.

b) In the same discussion, on p.12 of [Wiseman et al., 2022], WCR seem, at first reading, to define an "independent" party to be one whose "thoughts are not known to us unless he chooses to reveal them". "Not known to us" seems to depend on our technological, observational, and psychological powers, and potentially also on our level of scientific understanding of the relationship between thoughts and brain states. I would suggest that a metaphysical postulate (such as FRIENDLINESS) should not include terms (such as "independent") that have such dependences.

One might consider "Not knowable by us, regardless of our technology (etc.)" as an alternative. The problem with this is that it is not obvious – or even particularly plausible -- that any parties involved in a Wigner's Friend experiment whose outcomes we observe can be independent by this definition. It seems quite likely, for instance, that neuroscience and technology will eventually advance to the point where scans allow every human's thoughts to be reliably inferred by others. As far as I can see, the only clear examples of genuine independence, by this definition, are parties permanently spacelike separated from us.

Perhaps, though, WCR's intended definition is different from either of the above. The full discussion of independence on p.12 of [Wiseman et al., 2022] is:

> "Finally, for a party to be "independent", it is sufficient that his thoughts are not known to us unless he chooses to reveal them. That is, while this party's thoughts may (of course) be influenced by how other parties interact with him, his thoughts are not directly "implanted"."

Strictly speaking, the first sentence is a statement of sufficiency, not a definition; the sentence does not logically imply the condition is necessary. Confusingly, although the second sentence begins "That is", it makes a significantly different statement. Being able to infer someone's thoughts is different from being able to control them. I suspect that in fact the first sentence is misleading and it is the second one that is important, i.e., that WCR are really concerned with excluding thought control or implantation rather than mind-reading: the former makes more sense in the context of their arguments. I suspect too that the definition should be taken as applying to the state of the party during the open-ended communication in which they "[display] cognitive ability at least on par with my own", not to their state in all conceivable circumstances. That is, so long as his thoughts are not implanted or controlled by us during the relevant communication, it does not matter whether we have or use the power to do so at other times.

**Acknowledgements**

Many thanks to all the participants at the 2022 San Francisco "Wigner's Friends" workshop for very helpful, enjoyable and stimulating discussions and presentations, and in particular to Eric Cavalcanti, Eleanor Rieffel, Howard Wiseman, who also made very helpful comments on earlier drafts of this paper. Thanks too to Matthew Leifer and Mackenzie Dion for helpful discussions and to Mackenzie Dion for drawing my attention to relevant references.

This work was supported by FQXi and by Perimeter Institute for Theoretical Physics. Research at Perimeter Institute is supported by the Government of Canada through Industry Canada and by the Province of Ontario through the Ministry of Research and Innovation. I also gratefully acknowledge financial support from the UK Quantum Communications Hub grant no. EP/T001011/1.

**References**

**Valia Allori, Angelo Bassi, Detlef Dürr, and Nino Zanghi, editors. "Do wave functions jump?". Fundamental Theories of Physics. Springer. Switzerland (2021).**

**Adam B. Barrett, "An integration of integrated information theory with fundamental physics", Front. Psychol., Sec. Consciousness Research, 5 (2014) https://doi.org/10.3389/fpsyg.2014.00063**

**Kok-Wei Bong, Aníbal Utreras-Alarcón, Farzad Ghafari, Yeong-Cherng Liang, Nora**

Tischler, Eric G. Cavalcanti, Geoff J. Pryde, and Howard M. Wiseman. "A strong no-go theorem on the Wigner's Friend paradox". Nature Physics 16, 1199–1205 (2020).

Caslav Brukner. "A no-go theorem for observer-independent facts". Entropy 20, 50 (2018).

David Chalmers and Kelvin McQueen. "Consciousness and the collapse of the wave function." In "Consciousness and Quantum Mechanics", edited by Shan Gao, Oxford University Press, Oxford (2022).

Adrien Doerig, Aaron Schurger, Kathryn Hess, and Michael H Herzog. "The unfolding argument: Why IIT and other causal structure theories cannot explain consciousness". Consciousness and cognition, 72:49-59, (2019).

Hugh Everett. "'Relative state' formulation of quantum mechanics". Rev. Mod. Phys. 29, 454–462 (1957).

Daniela Frauchiger and Renato Renner. "Quantum theory cannot consistently describe the use of itself". Nature Communications 9, 3711 (2018).

Christopher A. Fuchs and Rüdiger Schack. "Quantum-Bayesian coherence". Rev. Mod. Phys. 85, 1693–1715 (2013).

Giancarlo C. Ghirardi, Alberto Rimini, and Tullio Weber. "Unified dynamics for microscopic and macroscopic systems". Phys. Rev. D 34, 470–491 (1986).

Marwan Haddara and Eric G. Cavalcanti. "A possibilistic no-go theorem on the Wigner's Friend paradox" arXiv:2205.12223 (2022).

Nicholas K. Humphrey. "Vision in a monkey without striate cortex: a case study." Perception 3.3 241-255 (1974).

Nicholas Humphrey. "Sentience: The Invention of Consciousness". Oxford University Press, Oxford, (2022).

Adrian Kent. "Causal quantum theory and the collapse locality loophole". Physical Review A 72, 012107 (2005).

Adrian Kent. "Quantum reality via late-time photodetection." Physical Review A 96 062121 (2017).

Adrian Kent. "Stronger tests of the collapse-locality loophole in Bell experiments". Phys. Rev. A 101, 012102 (2020).

Adrian Kent, "Collapse and Measures of Consciousness." Found Phys 51, 62 (2021)

Matthew Leifer, private communication (2022).

Janet Levin. "Functionalism", The Stanford Encyclopaedia of Philosophy (Winter 2021 Edition), Edward N. Zalta (ed.) (2021),

URL = <https://plato.stanford.edu/archives/win2021/entries/functionalism/>.

Johnjoe McFadden, "Integrating information in the brain's EM field: the cemi field theory of consciousness" Neuroscience of Consciousness, 1, niaa016 (2020), https://doi.org/10.1093/nc/niaa016

Markus Müller and Caroline Jones, "Why we should study other observer puzzles together with Wigner's Friend", talk at "Wigner's Friends" workshop, San Francisco (2022).

Masafumi Oizumi, Larissa Albantakis, and Giulio Tononi. "From the phenomenology to the mechanisms of consciousness: integrated information theory 3.0". PLoS Comput Biol, 10(5):e1003588, (2014).

Derek Parfit. "Reasons and Persons". Oxford University Press, Oxford, (1984).

Dimitris A. Pinotsis and Earl K. Miller, "Beyond dimension reduction: Stable electric fields emerge from and allow representational drift", NeuroImage 253, 119058 (2022).

Carlo Rovelli. "Relational quantum mechanics". International Journal of Theoretical Physics 35, 1637–1678 (1996).

Giulio Tononi. "Integrated information theory". Scholarpedia, 10(1):4164, (2015).

Giulio Tononi. "Consciousness as integrated information: a provisional manifesto". The Biological Bulletin, 215(3):216-242, (2008).

Naotsugu Tsuchiya, Thomas Andrillon, and Andrew Haun. "A reply to "The unfolding argument": Beyond functionalism/behaviorism and towards a truer science of causal structural theories of consciousness". PsyArXiv. June 7. doi:10.31234/osf.io/a2ms9, (2019).

David Wallace. "The emergent multiverse: Quantum theory according to the Everett interpretation". Oxford University Press, (2012).

Lawrence M. Ward and Ramón Guevara, "Qualia and Phenomenal Consciousness Arise From the Information Structure of an Electromagnetic Field in the Brain", Front. Hum. Neurosci., Sec. Cognitive Neuroscience, 16 (2022)
https://doi.org/10.3389/fnhum.2022.874241

Lawrence Weiskrantz. "Blindsight: A case study spanning 35 years and new developments." Oxford University Press, (2009).

Eugene P. Wigner. "Remarks on the mind-body question". In I. J. Good, editor, The Scientist Speculates. Heinemann, London (1961).

Howard M. Wiseman, Eric G. Cavalcanti, and Eleanor G. Rieffel, "A "thoughtful" Local Friendliness no-go theorem: a prospective experiment with new assumptions to suit", arXiv:2209.08491v1 (2022).